\documentclass[useAMS,usenatbib]{mn2e}
\usepackage{graphicx}
\usepackage{amssymb}
\usepackage{color}
\usepackage{natbib}
\usepackage{verbatim}

\def \aj {AJ}
\def \mnras {MNRAS}
\def \apj {ApJ}
\def \apjs {ApJS}

\def \aap {A\&A}

\def \prd {PhRvD}

\def \jcap {JCAP}
 
\title[Cosmological parameters from SN data]{Cosmological Parameter Estimation from SN~Ia data: \\a Model-Independent Approach} 
\author[S.Benitez-Herrera et al.]
{S.~Benitez-Herrera$^{1}$\thanks{E-mail:
benitez@mpa-garching.mpg.de}, E.E.O.~Ishida$^{1, 2}$, M.~Maturi$^{3}$, W.~Hillebrandt$^{1}$,
\newauthor
M.~Bartelmann$^{3}$, F.~R{\"o}pke$^{4}$\\
\vspace{0.5cm}
\\$^{1}$Max Planck Institute f\"ur Astrophysik, Karl-Schwarzschild-Str.~1,
D-85741 Garching, Germany
\\$^{2}$IAG, Universidade de S\~ao Paulo, Rua do Mat\~ao 1226, Cidade Universit\'aria, CEP 05508-900, S\~ao Paulo, SP, Brazil
\\$^{3}$Zentrum f\"ur Astronomie, ITA, Universit\"at Heidelberg, Albert-\"Uberle-Str.~2, 69120 Heidelberg, Germany
\\$^{4}$Universit{\"a}t W{\"u}rzburg, Emil-Fischer-Str.~31, D-97074 W{\"u}rzburg, Germany
}

\begin{document}

\date{Accepted 2013 August 23. Received 2013 August 22; in original form 2013 June 22}

\pagerange{\pageref{firstpage}--\pageref{lastpage}} \pubyear{2013}

\maketitle

\label{firstpage}

\begin{abstract}

\noindent We perform a model independent reconstruction of the cosmic expansion rate based on type Ia supernova data. Using the Union 2.1 data set, we show that the \textit{Hubble} parameter behaviour allowed by the data without making any hypothesis about cosmological model or underlying gravity theory is consistent with a flat $\Lambda$CDM universe having $H_0 = 70.43 \pm 0.33$ and \textbf{$\Omega_\mathrm{m}=0.297 \pm 0.020$,} weakly dependent on the choice of initial scatter matrix. This is in closer agreement with the recently released Planck results ($H_0 = 67.3 \pm 1.2, \Omega_\mathrm{m}  = 0.314 \pm 0.020$) than other standard analyses based on type Ia supernova data. We argue this might be an indication that, in order to tackle subtle deviations from the standard cosmological model present in type Ia supernova data, it is mandatory to go beyond parametrized approaches. 

\end{abstract}

\begin{keywords}
supernovae: general -- cosmology: cosmological parameters -- 
cosmology: observations -- cosmology: theory

\end{keywords}

\section{Introduction}

\noindent Supernovae Ia (SNe~Ia) are regarded as the best (relative) distance
indicators out to redshifts of $z \sim 1$. The first observational evidence of the 
accelerated cosmic expansion rate was provided by SNe~Ia
more than a decade ago \citep{riess1998a,perlmutter1999a}. 
Subsequently, it was supported by results from cosmic microwave background observations \citep{spergel2003} and baryon acoustic oscillations \citep{eisenstein2005}, that confirmed the presence of a component other than matter
in the Universe.
General relativity can accommodate the detected acceleration as an elastic and smooth fluid (dark energy) exerting a repulsive gravity 
\citep{turner1997a}. It fails, however, on giving a deeper understanding about its
cause. Other possibilities, such as quintessence models \citep[]{frieman1995a}
and non-standard cosmologies describing the acceleration as a manifestation
of new gravitational physics \citep[]{amendola2007a}, have been suggested as
alternative explanations for the acceleration. Hence, we face a situation where a large
variety of cosmological models has been proposed to account for cosmic
acceleration.
The standard approach of testing each model individually and determining its best fit parameters cannot account for unexpected features in the \textit{true} underlying cosmology and leads only to a smaller but still vast class of models allowed by the data.

Given the lack of a consistent physical framework to explain the energy component responsible for the cosmic acceleration, we aim at determining the dependence on the cosmic expansion with redshift (\textit{Hubble} parameter, \textit{H}) making use of only minimum hypotheses. 
The idea of a model-independent reconstruction extracted from distance measurements has been widely discussed in the literature. 
It was proposed by \citet{starobinsky1998a} and since then
other reconstructions of the kind have been carried out. \citet{shafieloo2006a} and \citet{shafieloo2007a} proposed
a procedure for smoothing supernova data over redshift with Gaussian kernels. \citet{fay2006a} added constraints from
measurements of baryon acoustic oscillations (BAO) to the SN~Ia data,
and \citet{daly2003a,daly2004a} combined SNe Ia
luminosity distances with angular-diameter distances from radio
galaxies. \citet*{seikel2008a,seikel2009a} have also tested the significance of
cosmic expansion directly from SN~Ia data in a model-independent way. 
Other non-parametric approaches to reconstruct the expansion history and equation of state of dark energy
use Gaussian Processes for smoothing the data \citep[see e.g.][]{seikel2012a}.

In this work, we use the method presented by \citet{mignone2008a} and further developed by \citet{maturi2009a}. It belongs to the class of purely geometrical approaches which only assumes a Friedman-Robertson-Walker metric, similar in philosophy  to the analysis presented by \citet{shafieloo2006a, ishida2011}. The method takes into account our expectations towards the underlying cosmology in order to define an optimal basis for the reconstruction, but this does not prevent it from being able to handle unexpected features that might be present in the data. At the same time, it provides a framework where subtle features can be addressed with a small number of free parameters and consequent reasonable uncertainties \citep[we refer the reader to][and references therein for a detailed analysis of the method]{maturi2009a,benitez2012}.  As expected from a completely geometrical approach, our method does not provide direct constraints on specific cosmological parameters. However, it returns an expected form of the expansion rate as a function of redshift with corresponding uncertainties. Having such a function, we can compare our results with $\Lambda$CDM cosmologies and point to the most likely cosmological parameters able to reproduce the resulting \textit{Hubble} parameter behaviour.

We have applied the method to the Union2.1 SNe Ia data set \citep{suzuki2012} and found a suitable reconstruction for \textit{H} using only one coefficient for the basis.
The \textit{Hubble} constant, $H_0$, is left as a free parameter to fit together with the expansion coefficients. 
It is worth mentioning that the SN data have to be standardized before extracting their absolute magnitudes, and subsequently distances. Several light-curve fitters available in literature approach the issue in different ways, their correction parameters having different physical meanings. The standardization process is a prior and separate step independent of the work presented in this paper. The data used here have been calibrated by the Union team \citep{kowalski2008a,suzuki2012} within the SALT2 (Spectral Adaptive Lightcurve Template) paradigm.
SALT2 \citep{guy2007a} is an empirical model based on a combination of photometric light curves and spectra of both nearby and distant SN~Ia. The photometry for each SN light curve is fitted to the model to determine a shape-luminosity parameter, a color parameter, and an overall flux normalization.
The global nuisance parameters are fitted simultaneously with the cosmological parameters while the \textit{Hubble} constant is marginalised over \citep[][]{suzuki2012}. The Union2.1 data set consists of a compilation of 580 SNe from several surveys providing redshift, distance moduli and errors in distance moduli.

Translating our results for a flat $\Lambda$CDM scenario leads to a cosmological constant model where matter energy density is in close agreement with the recently released Planck-satellite results for the cosmic microwave background \citep{planck2013}. Given the general agreement that the \textit{true} underlying cosmological model should not differ much from a cosmological-constant model (at least as long as the cosmic dynamics is concerned), we believe our results show that going beyond the parametrized analysis is fundamental to tackle small deviations present in the data.

This letter is organized as follows. In Section~\ref{sec:method}, the
essential aspects of the model-independent methodology are
reviewed. The application of the method to luminosity-distance
measurements and a comparison with $\Lambda$CDM models is presented in Section~\ref{sec:real data}. 
Finally, conclusions are drawn and future perspectives are discussed in Section ~\ref{conclusions}. 

\section{Model-independent method}
\label{sec:method}

\subsection{Recovering the expansion history of the Universe}
\label{sec:fitting}

We used the method presented in \citet[]{mignone2008a} and further developed in \citet[]{maturi2009a,benitez2012}. 
It aims at recovering the expansion rate in a model-independent fashion,
i.e. without reverting to any assumptions on the dynamics of the Universe. 
This is achieved by transforming the luminosity distance definition, assuming a Robertson-Walker metric, 

\begin{equation} 
  D_{\mathrm{L}}(a)=\frac{c}{H_{0}}\frac{1}{a}\int^{1}_{a} 
  \frac{dx}{x^2E(x)}\equiv\frac{c}{H_{0}}
  \frac{1}{a}\int^{1}_{a}\frac{dx}{x^2}e(x),        
\label{eq:lum_dist}                
\end{equation}

\noindent into a Volterra integral of the second kind. Here $e(x)\equiv E^{-1}(x)$ is the inverse of the expansion function. 
In order to do so the derivative of Eq.~(\ref{eq:lum_dist}) is taken with respect to the scale factor $a$.
Re-arranging terms we obtain

\begin{equation} 
  e(a)=-a^3D'_{\mathrm{L}}(a)+a\int^{a}_{1}\frac{dx}{x^2}e(x).
\label{eq:volterra}                        
\end{equation}

\noindent Note that, for the sake of simplicity, this expression is derived for a flat universe. 
This choice, however, does not affect the fundamental method and can be dropped
without change of principle if needed\footnote{The luminosity distance has been scaled by $c/H_0$ and $H_0$ is considered a free parameter whenever a reconstruction is confronted with the data.}.

Equation~(\ref{eq:volterra}) has proven to be uniquely solved in terms of a Neumann series \citep[][]{arfken1995a},

\begin{equation}
  e(a)=\sum^{\infty}_{i=0}a^ie_{i}(a).
\end{equation}
Consequently, in order to perform the reconstruction from noisy data we need a well behaved determination for $D_\mathrm{L}$ (Equation \ref{eq:lum_dist}) and its derivative (to evaluate Equation \ref{eq:volterra}).
Therefore, we need to first properly smooth the data by fitting an adequate function $D_{\mathrm{L}}(a)$ to the measurements
in a model-independent way. This is conveniently done through an expansion of the luminosity 
distance into a series of orthonormal functions,
\begin{equation}
D_{\mathrm{L}}(a)=\sum^{J}_{j=0}c_jp_j(a).
\label{eq:dl_first}
\end{equation}

The $J$ coefficients $c_j$ are those which minimize the $\chi^2$ statistic
function when fitting to the data. The number of terms to be included in the expansion depends 
strongly on the choice of the orthonormal basis and the quality of the data. 
Therefore, although the basis would be arbitrary with ideal data, it will not be in practice. 
In \citet{benitez2012}, we saw that by choosing a completely arbitrary basis, obtained
via Gram-Schmidt orthonormalization of the linearly-independent set $u_j=x^{(j/2-1)}$,
we needed at least three coefficients to fully reconstruct the expansion rate from Union2 data \citep{amanullah2010a}.
Despite this limitation, we were able to produce an acceptable fit of $H(a)$, although we
observed a systematic trend on its slope at intermediate and high redshifts 
when compared with $\Lambda$CDM or dark energy models.
This indicates that the estimation of the derivatives was not as accurate as one should expect. 
In this letter, we make use of an optimal basis system derived from a principal component analysis
which minimizes the number of coefficients required 
and orders them according to their information content.
It also removes any possible bias introduced by the choice of the basis. 

\subsection{Building the basis with Principal Component Analysis.}
\label{sec:basis}

\noindent Principal Component Analysis (PCA) is a well known statistical tool which aims at reducing the dimensionality of an initially very large parameter space. The algorithm looks for directions of maximum variance within the data and constructs an orthonormal basis representing directions (the principal components, PCs) of maximum clustering, or along which most of the information is contained. After the PCs are determined, the original data can be re-written as a linear combination of some PCs, usually  a number much smaller than the dimensionality of the original parameter space \citep[for a careful review see][]{jollife2002}.

Different approaches using PCA have already been proposed to reconstruct 
the dark energy equation of state $w$ \citep[e.g.][]{huterer2003,simpson2006,huterer2007}, the Hubble parameter \citep{ishida2011} and the cosmic star formation history \citep{ishida2011b}.
Here we follow the method described in \citet{maturi2009a} 
to obtain an optimal basis system for a given cosmological data set, 
in this case a luminosity-distance SNe catalog. 
We shall use the PCA approach to substitute the arbitrary orthonormal basis mentioned 
in section \ref{sec:fitting}, as proposed by \citet{maturi2009a}.
It is important to emphasise that, although we only present results of applying the method 
to SNe Ia data, it could be also used to analyse any other observable which delivers 
standard candle or standard ruler measurements (e.g. CMB or BAO). 
A general parameterization (with independent parameters regardless 
the underlying physical assumptions) could be achieved by considering the principal 
components as cosmological eigen-cosmologies. In this context, observations 
would ``excite'' (i.e. make visible) a given number of modes according to their accuracy. 

We start by defining a 1D vector which collects the redshift/scale factor values for which there is a luminosity distance measurement in our catalog, $\mathbf{x}$. The next step is to choose a group of models we believe spans the set of viable cosmologies. Suppose we chose initially $M$ different cosmologies: for each one of them, we calculate the luminosity distance at the values of scale factor in $\mathbf{x}$, producing for each model a vector, $\mathbf{t}_i$.
This ensemble of models, 
${\mathbf T} = (\mathbf{t}_1,\mathbf{t}_2, ..., \mathbf{t}_M)$, referred as the \textit{training set}, initializes the method. 
Each training vector $\mathbf{t}_i$ corresponds to a particular behaviour of the observable as a function of scale factor. 
The matrix $\mathbf{T}_{M\times n}$ represents a convolution of all our expectations towards the underlying cosmology and will act as a synthetic data set in order to determine an ideal orthonormal basis.

Once the training set of models is defined, 
we built the so called \textit{scatter matrix} $\mathbf{S}$, 
which contains the differences between each 
training vector and a given reference vector that defines the origin of the parameter space.
This reference model ($D_\mathrm{L}|_{\rm ref}$) may be any combination of models within $\mathbf{T}$, and is usually set to be the mean of the training set $\mathbf{\bar{t}} \equiv \langle t \rangle$.  A different choice will only be reflected in the final number of PCs used in the reconstruction. If the reference model was chosen wisely, maybe one PC shall be enough to account for the deviations from the reference model present in the data. Otherwise, we may need a larger number of components in order to achieve the same reconstruction. However, the choice of the reference model is arbitrary and does not affect the nature of the basis or the reconstruction of $D_\mathrm{L} $. 

The principal components are found by solving the usual eigenvalue problem $\mathbf{w}_i=\lambda_i\mathbf{S}\mathbf{w}_i$
where $\lambda_i$ and $\mathbf{w}_i$ are the eigenvalues and the eigenvectors, respectively. 
The linear transformation leading to these PCs is defined in such a way that
it concentrates in only a few features all the information (or variance) 
regarding the deviations of the models in the training set from the 
reference vector. The eigenvector with the largest eigenvalue corresponds to the direction of maximum variance (first PC). The second PC corresponds to the direction defined by the eigenvector with second largest eigenvalue and so on. 

An important issue when working with PCA is to determine how many PCs one should 
take into account \citep[][chapter 6]{jollife2002}. The number of PCs to be included in our reconstruction 
can be based on the cumulative percentage of total variance represented by a set of $L$ PCs,

\begin{equation} 
  t_L=\left(\frac{\sum^L_{i=1}\lambda_i}{\sum^{N_\mathrm{PC}}_{j=1}\lambda_j}\times100\right),
\label{eq:power_pcs}                        
\end{equation}

\noindent where $N_\mathrm{PC}$ is the total number of PCs and $L$ the number to be
included in the reconstruction. In this way, the question of how many principal components 
to use translates into what percentage of variance  we are willing to consider.

After constructing the orthonormal basis and deciding how many PCs to include in the final analysis ($L$), we express the corrections to the reference model as linear combinations of the first $L$ PCs, 
\begin{equation}
D_{\mathrm{L}}(a)\equiv D_\mathrm{L}|_{\rm ref} + \sum^{L}_{j=0}c_jw_j(a). 
\label{eq:rec_dl}
\end{equation}
Following what was done in the previous section, the final values for the coefficients $c_j$ are determined by confronting this expression for the luminosity distance with the data through a $\chi^2$ minimisation. In this step, the \textit{Hubble} constant $H_0$ is considered a free parameter to be minimised along with the coefficients $c_j$. 
Subsequently, we approximate the derivative in Eq.~(\ref{eq:volterra})  as
\begin{equation}
  D'_{\mathrm{L}}(a)=\sum^{L}_{j=0}c_jw'_j(a).
  \label{eq:deriv_dl}
\end{equation}
Due to the linearity of Eq.~(\ref{eq:volterra}), it is possible to
compute it for each mode $j$ of the basis separately.
Thus, the final solution in terms of Neumann series is
\begin{equation}
  e(a)=\sum^{L}_{j=0}c_je^{(j)}(a),
\end{equation}
\label{eq:final_ex}
that is, the measured coefficients give the
solution for the expansion function. 

It is also important to stress that the method is able to constrain other cosmologies 
which are not explicitly included in the original training set. This again can be achieved by using of a larger number of PCs \citep[see Section 4.2 of][]{maturi2009a}.

\subsection{Error analysis}

The errors in our method arises mainly from the uncertainty in the determination of the expansion coefficients ($c_j$) and $H_0$ 
(which is left as a free parameter) due to the minimisation. 

The errors in the coefficients then propagate into the estimate of $e(a)$ as follows
\begin{equation}
[\Delta e(a)]^2=\sum^J_{j=0}\left[\frac{\delta e(a)}{\delta c_j}\right]^2(\Delta c_j)^2=\sum^J_{j=0}\left[e^{(j)}(a)\right]^2(\Delta c_j)^2.
\end{equation}

\noindent The final errors on the expansion rate $E(a)=1/e(a)$ are

\begin{equation}
[\Delta E(a)]^2=\frac{[\Delta e(a)]^2}{e^4(a)}
\end{equation}

The error contribution due to the minimisation of $H_0$ is added to the previous one in quadrature.
Moreover, the uncertainty in our ability to determine the principal components
is given by $\sigma_{\mathrm{PC}_i}\sim\frac{1}{\sqrt\lambda_i}$ \citep{ishida2011, jollife2002}.
In this way, the total error budget is expressed as $\sigma_T^2=\sigma_{coeff}^2+\sigma_{H_0}^2+\sigma_{PC_i}^2$.

\section{Application to real data: beyond Union2}
\label{sec:real data}

We applied the method described above to the largest homogeneously reduced SN~Ia sample publicly available, the Union2.1 \citep{suzuki2012}. This sample contains 580 SNe and includes
data from SNLS \citep{astier2006a}, ESSENCE \citep{miknaitis2006a} and SDSS \citep{holtzman2008a}
surveys, low redshift samples \citep{hamuy1996a,hicken2009a} as well as \textit{Hubble Space Telescope} data \citep{riess2007a}.

\begin{figure}
\begin{center}
\includegraphics[height=70mm,width=92mm,clip=true]{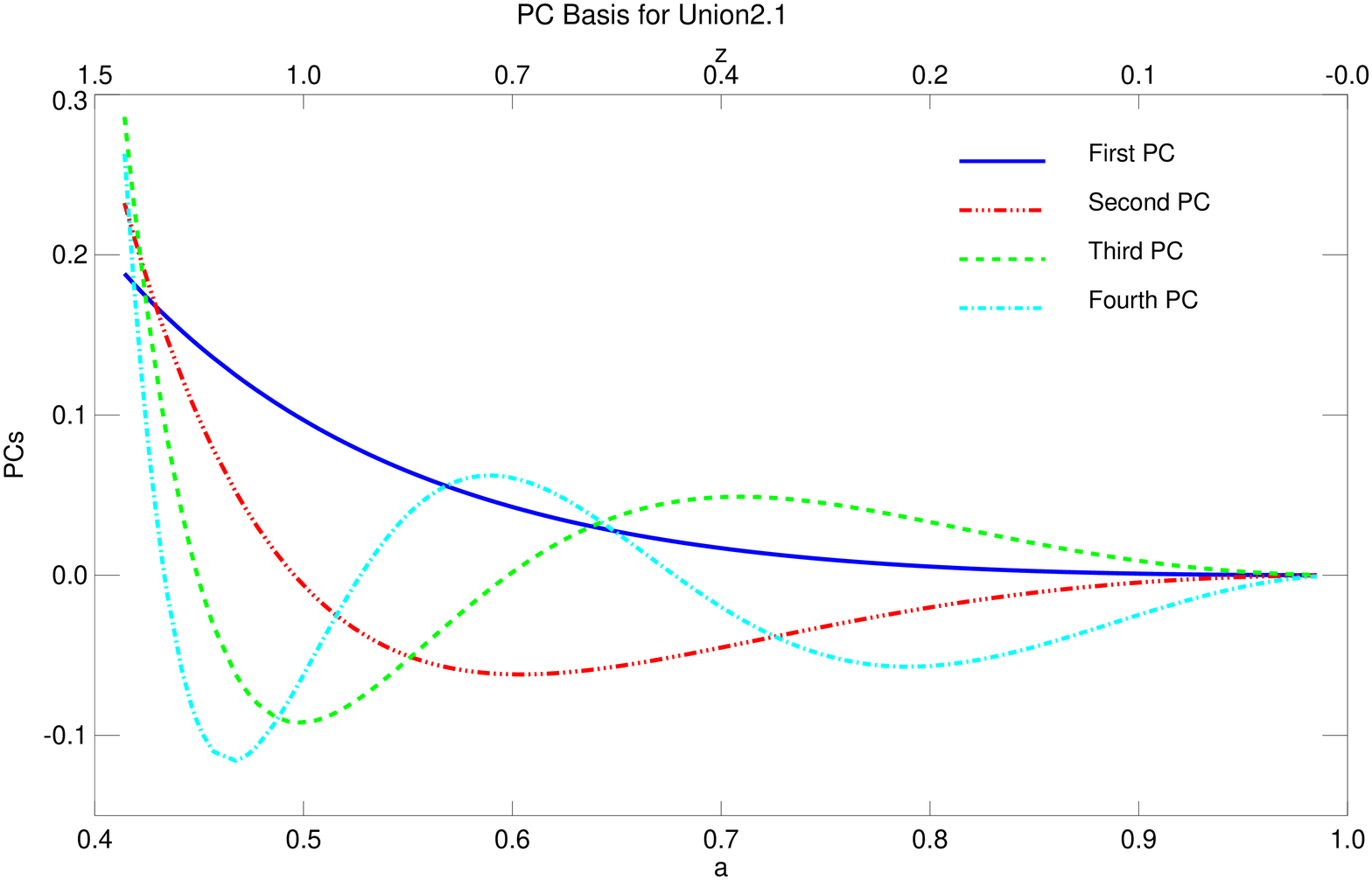} 
\caption{First four principal components built using the Union2.1 redshift coverage and a training set 
of $\Lambda$CDM models with the following sampling of the cosmological 
parameters: $0.1\leq\Omega_\mathrm{m}\leq0.5$ and $0.5\leq\Omega_{\Lambda}\leq0.9$.}
\label{fig:pcs}
\end{center} 
\end{figure}

In Fig.~\ref{fig:pcs} we show the
principal components obtained using the redshift coverage of Union2.1 and a training set 
of 11 flat $\Lambda$CDM models with the sampling $0.1\leq\Omega_\mathrm{m}\leq0.5$ and $0.5\leq\Omega_{\Lambda}\leq0.9$. 
The first principal component for this sample already accounts for $\approx99.9\%$ of the total variance and its determination carries an uncertainty of $\sigma_\mathrm{PC_1}\approx 2.5\times 10^{-5}$. This means it already contains the main properties of the expansion of the Universe and accounts alone for a great part of the total variance sampled in the scatter matrix. Therefore, we restrict ourselves to only one principal component when performing the reconstruction.

Fitting the luminosity distance data to the expression in Eq. (\ref{eq:rec_dl}) with 1 PC returns $c_1 = 122.74 \pm 796.68$ and $H_0 = 70.43 \pm 0.33$. These specific values were obtained considering $D_\mathrm{L}|_{\rm ref} = \mathbf{\bar{t}}$. However we did test other reference models and, although the value for $c_1$ depends slightly on the choice of $D_\mathrm{L}|_{\rm ref}$, the final reconstruction and the minimised $H_0$ do not (this is not the case in general and is due to the fact that the space covered by the training set is in a volume tightly enclosing the data). It is worth mentioning that the errors we obtain for $H_0$ are purely statistical and only due to the minimization process. They are negligible compared with the calibration error. However, we present our best fit here to demonstrate that the reconstructed zero-point of $H(z)$ points towards a lower value than the standard approach. This trend has been confirmed by the Planck results \citep{planck2013}.  

\begin{figure}
\includegraphics[height=70mm,width=92mm,clip=true]{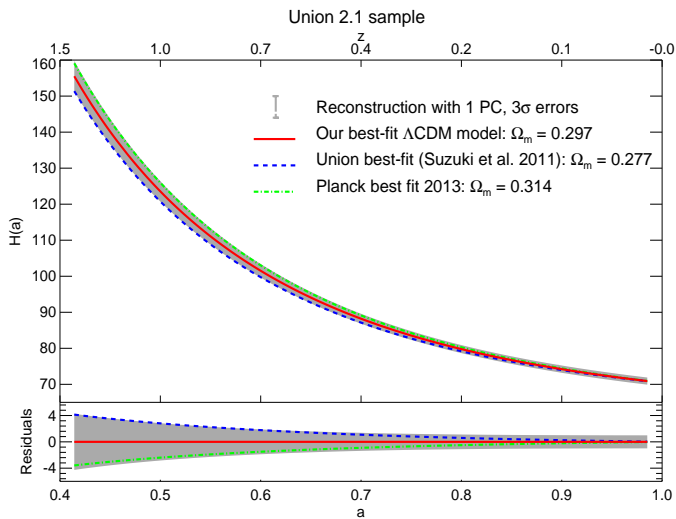}
\caption{The reconstructed expansion history, with $3\sigma$ error,
  extracted from the Union2.1 sample using the optimal basis system with one PC. 
The red (solid) line represents our best-fit to $\Lambda$CDM paradigm. The blue (dashed) line is the best-fit 
obtained by the Union team for this sample and the green (dot-dashed) line corresponds to results from Planck.}
\label{fig:Union2}
\end{figure}

The gray area in Fig. (\ref{fig:Union2}) represents the reconstructed expansion history within 3$\sigma$ errors. For the sake of comparison, the figure also shows the best-fit cosmology  found by the original Union2.1 analysis, $\Omega_\mathrm{m}=0.277\pm0.022$ \citep{suzuki2012}  and the latest result reported by the Planck satellite team, $\Omega_\mathrm{m}=0.314\pm0.020$ \citep{planck2013}. In order to avoid confusion, only the best-fit curves are shown in Fig. (\ref{fig:Union2}). Both results are in marginal agreement with the behaviour we found for $H(a)$.

For the reasons exposed above, it is clear that our approach does not output fits to specific cosmological parameters. 
However, we can put our reconstruction in the context 
of $\Lambda$CDM models and find the range of $\Omega_\mathrm{m}$ values allowed by the behaviour we found for $H(a)$. Keeping fixed the value we found for $H_0$, we obtain $\Omega_\mathrm{m}=0.297\pm0.020$ (red curve in Fig.~\ref{fig:Union2}). We emphasise that the magnitude of the error does not carry the same meaning as in the standard parametric analysis shown in Fig. \ref{fig:Union2}. The determination of range of values for $\Omega_\mathrm{m}$ is merely a strategy to better compare our results. Unlike other analyses we are tracking only dynamical deviations from $\mathbf{\bar{t}}$, which causes the error bars to be small. 

\begin{figure}
\includegraphics[height=70mm,width=92mm,clip=true]{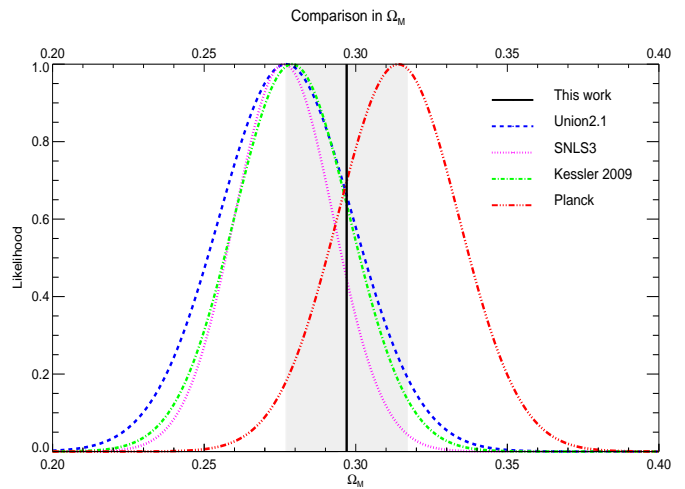}
\caption{Comparison between our results when translated to flat $\Lambda$CDM scenario, $\Omega_\mathrm{m}=0.297\pm0.020$ (gray region), and others from the literature. The green (dot-dashed) line correspond to results reported by the first year of SDSS data, $\Omega_\mathrm{m}=0.279\pm0.019$ \citep{kessler2009b}. Cyan (dashed)  line stands for results reported by Union2.1 team, $\Omega_\mathrm{m}=0.277\pm0.022$ \citep{suzuki2012}, the pink (dotted) line represents outcomes from SNLS3, $\Omega_\mathrm{m}=0.276\pm0.016$ \citep{sullivan2011a} and the red (dot-dot-dashed) line represents recent results from Planck, $\Omega_\mathrm{m}=0.314\pm0.020$ \citep{planck2013}. Only statistical errors are considered in this plot.}
\label{fig:gaussian}
\end{figure}

The results we found are significantly higher than the best-fit value obtained by the Union2.1 team ($\Omega_\mathrm{m}=0.277$ without systematics; blue curve in Fig.~\ref{fig:Union2}), and are in close agreement with the value reported by  Planck. We show in Fig. (\ref{fig:gaussian}) a more clear comparison of our results with others from the SNe Ia literature \citep[all using SALT2 light curve fitter;][]{guy2007a}. We believe that the shift in our results towards the Planck values is an indication that SNe Ia cosmology should move beyond the parametrized approaches if it aims at dealing with small deviations from the standard values of the cosmological parameters present in the data. We want to stress again that we are not biased by any theoretical opinion towards a cosmological model, since we do not assume any specific form of the Friedmann equations. In fact, the method is a powerful tool to evaluate non-standard cosmologies as we tentatively showed in \citet[][]{benitez2012} and plan to explore in a more rigorous way in future work. 

Our model-independent method to constrain the expansion history
has other interesting applications. 
For instance, it offers a complementary way of detecting possible systematic
effects -- e.g. the uncertainties introduced by the light curve calibration, 
host galaxy extinction or intrinsic variations corresponding to different SN Ia populations --
which could affect the data and would be overlooked within a
traditional analysis based on physical
parameterizations. The method is also a valuable tool that can be used to plan
future Type Ia supernova cosmology campaigns, by testing 
redshift ranges in which it would be more relevant to collect data \citep[see][]{benitez2012}.

\section{Conclusions}
\label{conclusions}

\noindent In this letter, we have applied a method to recover the expansion 
history of the Universe in a model-independent fashion.
The luminosity-distance
measurements, obtained from SNe~Ia, depend only on space-time
geometry, and can be directly related to the \textit{Hubble} function without
assuming any dynamical model.
We argue that, as long as the nature of dark energy remains unknown, 
model-independent analyses of the kind described here
have more significance in deriving cosmological parameters 
than traditional parametric studies. 
This is due to the fact that no specific form of the equation 
of state $w$ or the \textit{Hubble} function is fixed in our approach. 
Our only assumption relies on the argument that the luminosity 
distance can be expanded into a series of orthonormal functions.
This basis is chosen to be derived from PCA, in an attempt 
to control the number of coefficients to be included in our reconstruction
in a rigorous way.

Our analysis shows that SNe data do point to a higher value of $\Omega_\mathrm{m}$, 
in contradiction with what was found with standard methods. 
Furthermore, this is in agreement with the last results driven by Planck 
and might be an indication that it is important to move beyond parametric fits.

The ultimate goal of this work is to discriminate among different cosmological models,
such as $f(R)$ or DGP theories, based on very different physical assumptions, 
and, in this way, break the current degeneracy in the cosmological parameters.
Further analysis on simulated data might point to caveats not appearing in real data analysis.

The model-independent method 
offers a complementary way of detecting possible systematic errors.
This is especially relevant if one considers the dependence introduced by 
the light-curve fitters on the derived distances moduli from SNe~Ia.
It is important to test the performance of the available light-curve fitters
on the base of model-independent approaches. 
A similar analysis of the same data set with different light curve fitters is important 
and will certainly be presented in a subsequent work.

Finally, is it worthwhile noting
the potential of the method for the analysis of possible local
inhomogeneities through comparison of the expansion history in
different directions.

\section*{Acknowledgments}
It is a pleasure to thank A. Shafieloo and A. Kim for providing helpful comments and suggestions. 
E.E.O.I. thanks the Brazilian agency FAPESP (2011/09525-3) for financial support.
This work was also partially supported by the Deutsche Forschungsgemeinschaft via the Transregional 
Collaborative Research Center ``The Dark Universe'' (TRR~33), the Emmy Noether Program (RO
3676/1-1), and the Excellence Cluster ``Origin and Structure of the
Universe'' (EXC~153).

\bibliographystyle{mn2e}


\end{document}